\documentstyle[12pt]{report}
\begin{document}


\pagestyle{empty}

\renewcommand{\thefootnote}{\fnsymbol{footnote}}


\begin{flushright}
{\small
SLAC--PUB--7174\\
hep-th/9605119\\
May 1996\\}
\end{flushright}

\begin{center}
{\bf\large   
Models of Dynamical Supersymmetry Breaking from a SU(2k+3) Gauge Model\footnote{Work supported by
Department of Energy contract  DE--AC03--76SF00515.}}

\smallskip

Chih-Lung Chou\\
Stanford Linear Accelerator Center\\
Stanford University, Stanford, California 94309\\
and\\
Applied Physics Department\\
Stanford University, Stanford, California 94309\\
alexchou@leland.stanford.edu\\
\medskip

\end{center}

\vfill

\begin{center}
{\bf\large   
Abstract }
\end{center}

\indent We investigate three classes of supersymmetric models which can be obtained by breaking the chiral SU(2k+3) gauge theories with one antisymmetric tensor and 2k-1 antifundamentals. For N=3, the chiral SU(2k)$\times$SU(3)$\times$U(1) theories break 
supersymmetry by the quantum deformations of the moduli spaces in the strong SU(2k) gauge coupling limit. For N=2, it is the generalization of the SU(5)$\times$U(2)$\times$U(1) model mentioned in the literature. Supersymmetry is broken by carefully choosi
ng the quark-antiquark-doublet Yukawa couplings in this model. For N=1, this becomes the well-known model discussed in the literature.

\vfill

\begin{center}

Submitted to {\it Physics Letters B}

\end{center}

\vfill\eject
\pagestyle{plain}
\pagenumbering{arabic}
{\LARGE{\noindent 1    Introduction}}
\vspace{0.5cm}

\indent Supersymmetry is considered to be a leading candidate for resolving the gauge hierarchy problem.  It may provide explanations of why the electroweak scale is much smaller than the Planck scale if supersymmetry is dynamically broken within a few or
ders of magnitude of the weak scale. Supersymmetric gauge theories also have the merit of providing some helpful aspects [1] to the understanding of the non-supersymmetric gauge theories.
 
\indent There are various models of dynamical supersymmetry breaking.  Many of them possess flat directions in field spaces in which the energies vanish classically.  The non-perturbative gauge dynamics then lead to SUSY breaking by generating a non-pertu
rbative superpotential which lifts the classical moduli spaces.  On the other hand, the gauge dynamics could lead to degeneracy of the quantum moduli space rather than generation of the non-perturbative superpotential in some SUSY theories.  A recent repo
rt by Intriligator and Thomas [2] observed that supersymmetry can also be broken by the deformation of the quantum moduli space.  This occurs if the quantum deformed constraint is inconsistent with a stationary superpotential.

\indent In this paper we discuss the low-energy effective theories with the heavy degrees of freedom being integrated out at the strong gauge scale.  The low-energy theory is described by the K$\ddot a$hler potential, the superpotential and the gauge coup
ling function with the latter two holomorphic to the chiral fields. The holomorphicity of the superpotential combined with the global symmetries then can be used to determine the exact effective superpotential [1].  If one is only interested to know that 
supersymmetry is broken then the knowledge of the K$\ddot a$hler potential is not necessary provided that the K$\ddot a$hler potential is non-singular.  The K$\ddot a$hler potential is necessary only in calculating the vacuum energy and the spectrum [3].

\indent The SU(2k+3) theory is well known to break supersymmetry once appropriate Yukawa terms are added to the superpotential [4a,4b].  It has a zero Witten index [5], which implies that if the SU(2k+3) theory undergoes symmetry breaking, the resulting S
U(2k+3-N)$\times$SU(N)$\times$U(1) theory has a zero Witten index as well.  This makes the SU(2k+3-N)$\times$SU(N)$\times$U(1) theory
a good candidate for Supersymmetry breaking. In the following sections, we will discuss those theories with  N$\leq$3 in the limit where the SU(N) coupling is much weaker at the scale at which SU(2k) becomes strong. 

\indent This paper is organized as follows. In section two we investigate the SU(2k)$\times$SU(3)$\times$U(1) models.  Under the strong SU(2k) gauge coupling assumption, the SU(2k) gauge dynamics deformed the quantum moduli space instead of generating a n
on-perturbative term in the effective superpotential.  There are extra SU(3) triplets in this model as compared to the pure SU(2k) gauge model [4].  It will be shown that these triplets play crucial roles in breaking supersymmetry.

\indent In section three, we discuss the resulting SU(2k+1)$\times$SU(2)$\times$U(1) gauge theories which were first proposed by M. Dine and his colleagues [6] and a SU(5)$\times$SU(2)$\times$U(1) model was discussed and shown to break SUSY in their paper
. These models do have R-symmetry. All its flat directions are shown to be lifted once appropriate Yukawa coupling terms are added to the superpotential. For K=1, it becomes the conventional SU(3)$\times$SU(2) theory  except for an extra singlet field S.
  
\indent In section four we discuss the resulting SU(2k+2)$\times$U(1) model.  This model is already well know in the literature [4,6] with SUSY being dynamically broken.  We give a brief summary of the published result in this section.  

\indent We conclude the results in section five.

\vspace{1cm}

{\Large{\noindent 2. SU(2K)$\times$SU(3)$\times$U(1)}}

\vspace{0.6cm}

Recently,  Intriligator and Thomas [2] observed that SUSY can be broken by the deformation of the quantum moduli space in the SU(3)$\times$SU(2) model in the limit $\Lambda_2\gg\Lambda_3$.  This can be generalized to SU(2k)$\times$SU(3)$\times$U(1) theory
.  This model contains one antisymmetric tensor $A_{\alpha \beta}$ with U(1) charge -2, one quark $Q^a_{\alpha}(2k,3,-1+{2k \over 3})$, 2k-1 antiquark ${\bar Q}^{\alpha}_i({\overline {2k}}, 1,1)$, one SU(3) triplet $\bar f_a(1, {\bar 3}, {4k \over 3})$ an
d 2k-1 SU(3) triplets ${\bar F}_{ia} (1, {\bar 3},{-2k \over 3})$.  The SU(2k) invariant operators are defined by

\vspace{0.4cm}

\hspace{1cm}$M^a_i \equiv {\bar Q}^{\alpha}_i Q^a_{\alpha}$ 

\vspace{0.2cm}

\hspace{1cm}$X_{ij} \equiv A_{\alpha \beta} {\bar Q}^{\alpha}_i {\bar Q}^{\beta}_j$ 

\vspace{0.2cm}

\hspace{1cm}$Y_a \equiv \varepsilon_{abc} Q^b_{\alpha_1} Q^c_{\alpha_2} \varepsilon ^{\alpha_1 \cdots \alpha_{2k}} A_{\alpha_3 \alpha_4} \cdots  A_{\alpha_{2k-1} \alpha_{2k}}$ \hfill (1)

\vspace{0.2cm}

\hspace{1cm}$PfA \equiv \varepsilon^{\alpha_1 \cdots \alpha_{2k}} A_{\alpha_1 \alpha_2} \cdots  A_{\alpha_{2k-1} \alpha_{2k}}$

\vspace{0.4cm}

\noindent where $a=1 \cdots 3$,  $\alpha = 1 \cdots 2k$ and $i,j = 1 \cdots 2k-1$.  The D-flat directions are parameterized by the following gauge invariant chiral polynomials:

\vspace{0.3cm}

\hspace{1cm} \makebox[3.5cm][l]{$X_{ij},$} \makebox[3.5cm][l]{$C_{ijk} \equiv (M_i \cdot M_j \cdot M_k) PfA$}

\vspace{0.2cm}

\hspace{1cm} \makebox[3.5cm][l]{$W_i \equiv Y M_i$,} \makebox[5cm][l]{$V_i \equiv ({\bar f} M_i)PfA,$}\makebox[3.5cm][l]{$U_{ij} \equiv {\bar F_i} M_j$}

\vspace{0.2cm}

\hspace{1cm} \makebox[3.5cm][l]{$B_i \equiv Y \cdot {\bar f} {\bar F_i}$}

\vspace{0.2cm}

\hspace{1cm} \makebox[3.5cm][l]{$E_{ij} \equiv {\bar f} \cdot {\bar F_i} \cdot {\bar F_j}$} \hfill (2)

\vspace{0.3cm}

\noindent where the dot multiplication are defined as $M_i \cdot M_j \cdot M_k = \varepsilon_{abc}M^a_iM^b_jM^c_k$.  These gauge invariant chiral polynomials are not all independent but subject to constraints.  In the limit where the SU(3) gauge coupling 
is weak at the scale $\Lambda_{2k}$, the non-perturbative SU(2k) gauge dynamics deforms the quantum moduli space[4] instead of generating an effective superpotential term.

\vspace{0.7cm}

$W_i\varepsilon^{i_1 \cdots i_{2k-1}}X_{i_2i_3} \cdots X_{i_{2k-2}i_{2k-1}} - $

\vspace{0.4cm}

\hspace{2cm} ${k-1 \over 3k} C_{i_1i_2i_3}\varepsilon^{i_1 \cdots i_{2k-1}}X_{i_4i_5}\cdots X_{i_{2k-2}i_{2k-1}} - \Lambda^{4k}_{2k} = 0$ \hfill (3)

\vspace{0.6cm}

\noindent This quantum modified constraint can be implemented by adding $W_{constraint}$ to the superpotential via a Lagrange multiplier field $\cal L$ 

\vspace{0.8cm}

$W_{tree} = \lambda^{ij}_1U_{ij} +  \lambda^{ij}_2X_{ij} + \lambda^{ij}_3 E_{ij}$.

\vspace{0.3cm}

$W_{eff} = \lambda^{ij}_1U_{ij} +  \lambda^{ij}_2X_{ij} + \lambda^{ij}_3 E_{ij}+{{\cal L} \lbrace W_{i1}\varepsilon^{i_1\cdots i_{2k-1}}X_{i_2i_3} \cdots X_{i_{2k-2}i_{2k-1}}} $

\vspace{0.3cm}

\hspace{2cm} $ - {k-1\over 3k}C_{i_1i_2i_3}\epsilon^{i_1 \cdots i_{2k-1}} \cdots X_{i_{2k-2}i_{2k-1}} - \Lambda^{4k}_{2k}\rbrace ,$ \hfill (4)

\vspace{0.8cm}

\noindent Here we choose to add renormalizable operators to $W_{eff}$ that preserve the non-anomalous $U(1)_R$ global symmetry. Without loss of generality the coupling matrix $\lambda_1$ can be assigned  a diagonal matrix.  It is also easy to see that not
 all $E_{ij}$ terms are allowed in the tree superpotential. This is because if all $E_{ij}$ terms are allowed then all $U(1)_R$ charges $R(\bar F_i)$ of $\bar F_i$ must be equal which is not consistent with both the anomaly free conditions and the $R(W_{t
ree})=2$ requirement. However, it is still very possible for the tree superpotential to have flat directions after fulfilling the above conditions.  To assign the Yukawa coupling matrices that result in no flat direction we need more information about the
 flatness conditions.

\vspace{0.5cm}

$\lambda_2^{ij} {\bar Q_i^{\alpha}}{\bar Q_j^{\beta}} = 0 $ \hfill (5.1)

\vspace{0.3cm}

$\lambda_3^{ij} \varepsilon^{abc} {\bar F_{ia}} {\bar F_{jc}} = 0$ \hfill (5.2)

\vspace{0.3cm}

$\lambda_1^{ij} {\bar F_{ia}} {\bar Q^a_j} = 0$  \hfill (5.3)

\vspace{0.3cm}

$\lambda_1^{ij} {\bar F_{ia}} Q^a_{\alpha} + 2 \lambda_2^{ij} A_{\alpha \beta} {\bar Q^{\beta}_i} = 0 $ \hfill (5.4)

\vspace{0.3cm}

$\lambda_1^{ij} {\bar Q^{\alpha}_j} Q^a_{\alpha} - 2 \lambda_3^{ij} \varepsilon^{abc} {\bar f_b} {\bar F_{jc}} = 0$ \hfill (5.5)

\vspace{0.5cm}
\noindent For simplicity, we may just assign the $\lambda_1$ matrix to be the unit matrix without loss of generality.  To reduce the complexity
it is observed that if the Yukawa couplings $\lambda_2$ and $\lambda_3$ are of the following forms, 
\vspace {0.3cm}

$$ |\lambda_2^{ij}| =\cases{nonzero& if i=f(j) and j=f(i)   \cr
			          0& otherwise   \cr}$$         
\vspace{0.3cm}

$$|\lambda_3^{ij}| =\cases{nonzero& if i=g(j) and j=g(i)  \cr
			          0& otherwise. \cr}$$ 

\noindent where f and g are one-to-one mappings of indices i and j, then equations (5.4) and (5.5) imply that operators $U_{ij}$ will have one-to-one correspondences of operators X and E respectively.

\vspace{0.4cm}

$U_{jk}$ = $-2\lambda^{f(j)j}_2X_{f(j) k}$, (no summation over j) \hfill (6.1)

\vspace{0.3cm}

$U_{kj} = 2\lambda^{jg(j)}_3E_{g(j)k}$,  (no summation over j) \hfill (6.2)

\vspace{0.3cm}

$U_{g(i),i}$ = $U_{i,f(j)}$ =0.  (no summation over i) \hfill (6.3)

\vspace{0.4cm}

\noindent By multiplying operators ${\bar f_b}Q^b_{\alpha_2} {\varepsilon^{\alpha \alpha_2 \cdots \alpha_{2k}}}A_{\alpha_3 \alpha_4} \cdots A_{\alpha_{2k-1} \alpha_{2k}}$ on both sides of the equation (5.4) we find:

\vspace{0.4cm}

$B_j + 2\lambda^{f(j)j}_2V_{f(j)} = 0$. (No summation over j) \hfill (7)

\vspace{0.3cm}

\noindent From equations (6) and considering the antisymmetry of $\lambda_2$ and $\lambda_3$ we get 

\vspace{0.4cm}

$U_{jk}= -{\lambda^{kg(k)}_3 \over \lambda^{g^{-1}(j)j}} U_{g(k), g^{-1}(j)} = -{\lambda_2^{f(j)j} \over \lambda_2^{kf^{-1}(k)}}U_{f^{-1}(k), f(j)} $. (No summation over j,k) \hfill (8)

\vspace{0.4cm}

\noindent Equation (8) is very powerful in forcing operators to have vanishing vev's.  To be more explicit, we give the flows of the indices

\vspace{0.4cm}

\noindent $(j,k)  \longrightarrow (g(k), g^{-1}(j)) \longrightarrow (f^{-1}g^{-1}(j), fg(k)) \longrightarrow (gfg(k), g^{-1}f^{-1}g^{-1}(j)) \longrightarrow  \cdots \cdots $

\vspace{0.3cm}

\noindent $(j,k) \longrightarrow (f^{-1}(k), f(j)) \longrightarrow (gf(j), g^{-1}f^{-1}(k)) \longrightarrow (f^{-1}g^{-1}f^{-1}(k), fgf(j)) \longrightarrow \cdots \cdots $.

\vspace{0.3cm}

\noindent If any one pair of indices in the follow represents a operator with vanishing vev then all other operators in the same chain will have vanishing vev's as well.  One possible assignment of $\lambda$ matrices that lifts all flat direction is given
 by Randall et. al.[8]

\vspace{0.5cm}

$ \lbrace(i, f(i))\rbrace = \lbrace (1,2), (3,4), \cdots , (2k-3, 2k-2) \rbrace$

\vspace{0.3cm}

$\lbrace(i, g(i))\rbrace = \lbrace (2,3), (3,4),  \cdots, (2k-2, 1)  \rbrace $.

\vspace{0.3cm}

\noindent Here the ordered pairs represent the unity entities of $\lambda_2$ and $\lambda_3$ matrices. Given the above assignments, the equations (5.4) and (5.5) lead to

\vspace{0.4cm}

$\bar Q^{\alpha}_{2k-1} Q^a_{\alpha} = 0 = M^a_{2k-1} = V_{2k-1} = U_{i, 2k-1} = W_{2k-1} = C_{( \cdots 2k-1 \cdots)}$

\vspace{0.3cm}

${\bar F_{2k-1,a}} Q^a_{\alpha} = 0 = U_{2k-1,j} = B_{2k-1}$ \hfill (9)

\vspace{0.3cm}

\noindent The equations (5.3) and (5.5) give

\vspace{0.4cm}

$W_i = 2 B_{g(i)}$

\vspace{0.3cm}

$V_i$ = 0,  $\rightarrow$ $B_i$ = 0 = $W_i$ \hfill for all i.  (10)

\vspace{0.3cm}

$U_{11}=U_{22}= \cdots U_{2k-2,2k-2} = 0$

\vspace{0.4cm}
                  
\noindent  To see explicitly why all operators have vanishing vev's we may take the case 2k=10 as an example.  In this case, the operators $U_{ij}$ are determined by using equations (6.1 $\sim$ 10).

\vspace{0.5cm}

$$U_{ij}=\left[\matrix{0&0&x&0&y&0&z&0&0   \cr
                       0&0&0&z&0&y&0&x&0    \cr
	               z&0&0&0&x&0&y&0&0    \cr	
                       0&x&0&0&0&z&0&y&0    \cr
                       y&0&z&0&0&0&x&0&0    \cr
	               0&y&0&x&0&0&0&z&0    \cr
                       x&0&y&0&z&0&0&0&0     \cr		 					
                       0&z&0&y&0&x&0&0&0    \cr
                       0&0&0&0&0&0&0&0&0    
 					}\right].$$       

\vspace{0.5cm}

\noindent Note that the symbols x, y and z used here are denoting the vevs of the same chains. If at least one of x, y and z is nonzero, say x $\not=$ 0, then we can always choose

\vspace{0.4cm}

${(\bar F_1)}_a = (t_{11}, 0, 0)$ \hfill  $t_{11}$ $\not=$ 0.

\vspace{0.3cm}

${(\bar F_2)}_a = (t_{21}, t_{22}, 0)$

\vspace{0.3cm}

\noindent and from $U_{11}, U_{12}$ and $U{13}$

\vspace{0.3cm}

${(M_1)} = (0, m_{12}, m_{13})$,  \hspace{0.8cm} ${(M_2)} = (0, m_{22}, m_{23})$

\vspace{0.3cm}

${(M_3)} = (m_{31}, m_{32}, m_{33})$

\vspace{0.4cm}

\noindent From $U_{21}$ and $U_{22}$ we know that $t_{22}$ $\not=$ 0 implies $m_{32}=m_{22}=0$ and thus contradicts the pattern that $U_{41}=0, U_{42}=x$$\not=$0.  If $t_{22}$ is zero then the vector $\bar F_2$ is parallel to the vector $\bar F_1$ if $t_{
21}$ is also nonzero.  However $U_{13}=x$ and $U_{23}=0$ tells us the impossibility for both $t_{11}$ and $t_{21}$ to be nonzero.  Therefore we conclude that the only possible way to satisfy the pattern is to put zero vev's in all the entries of the $U_{i
j}$ matrix.  Now we are left with the operators $C_{ijl}$.  That all the $C_{ijl}$ also have vanishing vev's can be seen as follows: to make all the $U_{ij}$ have vanishing vev's we could either set all vev's of ${\bar F_i}$ operators to zero or have at l
east one of them be nonzero.  By using equation (15.5) it is easy to check that all ${\bar F_i}$ having vanishing vev's implies the vanishing vev's of all $M_i$ operators.  On the other hand, if any one of the ${\bar F_i}$ operators has nonzero vev then a
ll of the $M_i$ should be of the following form:

\vspace{0.3cm}

$M_i = (0, m_{i2}, m_{i3})$.

\vspace{0.3cm}

\noindent Now we see that all $M_i$ lie on the same plane in a three-dimensional space and thus make any product $M_i \cdot M_j \cdot M_l$ to zero.  Therefore all of the $C_{ijl}$ are also lifted. 

Note that the nonzero entities of coupling matrices $\lambda_2$ and $\lambda_3$ are not necessary to equal to each other in magnitude.  The argument after equation (9) is also valid for nonzero coupling matrix entities having different numerical values.  
In other words, only the index-chain relation is important in breaking SUSY in these models. To form index chains, only even numbers of indices will be used in forming index pairs.  On the other side, in order to lessen the total numbers of chains (usuall
y means less unknown nonzero variables after determining $U_{ij}$ as that in the 2k=10 case) and lift more flat directions we should use as many indices as we can in forming index pairs.  Therefore k-1 index pairs are formed in each of the mappings f and 
g.

Given the tree level superpotential as above, the classical equations of motion could force all gauge invariant operators to have zero vev's thus lift all flat directions in the tree level superpotential.  On the other hand, the quantum constraint (3) for
ces some vev's to be non-zero which spontaneously breaks the $U(1)_R$ symmetry.  Therefore, the following SUSY-breaking criteria are satisfied[7]: 
  
\vspace{0.4cm}
\hspace{2cm}(1)There is no classical flat direction.

\vspace{0.3cm}
\hspace{2cm}(2)The global $U(1)_R$ symmetry is broken spontaneously.

\vspace{0.4cm}

\noindent According to the above criteria we can see that SUSY is likely broken dynamically.

In this section, we presume an non-anomalous $U(1)_R$ global symmetry and carefully select renormalizable operators for the tree level superpotential.  We see that choosing the Yukawa coupling matrices $\lambda_2^{ij}$ and $\lambda_3^{ij}$ is important in
 lifting classically flat directions.  Not all $U_{ij}$ and $E_{ij}$ terms can be chosen to preserve the non-anomalous global $U(1)_R$ symmetry. If we make one-to-one correspondences between operators $U_{ij}$ and $X_{ij}$ and $E_{ij}$ respectively, we ca
n preset some zero vev's in the $U_{ij}$ matrix by considering the antisymmetric characteristics of $X_{ij}$ and $E_{ij}$. The chain relations between index pairs of $\lambda_2$ and $\lambda_3$ then help to force all possible nonzero entities of $U_{ij}$ 
to zero.   

\vspace{1cm}

{\Large{\noindent 3.   SU(2K+1)$\times$SU(2)$\times$U(1)}}

\vspace{0.5cm}

\indent In this section we consider the SU(2k+3) gauge theory with one antisymmetric tensor and 2k-1 antifundamentals breaking down to SU(2k+1)$\times$SU(2) $\times$U(1).  The resulting model contains one antisymmetric tensor field $(A_{\alpha\beta})_{-2}
$, one fundamental $Q^{a}_{\alpha}(2k+1,2)_{-1+(2k+1)/2}$, 2k-1 antifundamentals $\bar Q^{\alpha}_i(\overline {2k+1},1)_{1}$, 2k-1 doublets $L^{a}_{i}(1,2)_{-(2k+1)/2}$ and one singlet field $S(1,1)_{2k+1}$.  The numbers on the lower right of parentheses 
represent charges of the U(1) gauge symmetry.  For k=1, except for the additional singlet field S, this is the well known SU(3)$\times$SU(2) model that has been investigated in the literatures and found to break SUSY either by the dynamically generated su
perpotential in the limit $\Lambda_3 \gg \Lambda_2$ or by the quantum deformation of the moduli space in the limit $\Lambda_2 \gg \Lambda_3$. For simplicity, we consider only theories where the SU(2) and U(1) gauge couplings are weak at the scale at which
 the SU(2k+1) subgroup becomes strong.  The SU(2k+1)$\times$SU(2)$\times$U(1) gauge invariant operators are given by

\vspace{0.2cm}

\hspace{1in} $\lbrace(S)_{2k+1},(C_{ij})_{2k+1}\rbrace \times \lbrace(L_{ij})_{-(2k+1)},(P_i)_{-(2k+1)}\rbrace$, 

\vspace{0.2cm}

\hspace{1in} $(N_{ij})_0\equiv L_i \cdot M_j$, \hspace{0.6cm} $(X_{ij})_0$,

\vspace{0.2cm}

\hspace{1in} $(W_i)_0\equiv Y \cdot M_i$. \hfill (11)

\vspace{0.3cm}

\noindent where operators $C_{ij}$, $L_{ij}$ and $P_i$ are defined by

\vspace{0.2cm}

$(C_{ij})_{2k+1}\equiv M_i \cdot M_j$ \hspace{0.7in} $(L_{ij})_{-(2k+1)}\equiv L_i \cdot L_j$, 

\vspace{0.2cm}

$(P_i)_{-(2k+1)}\equiv Y \cdot L_i$, \hfill (12)

\vspace{0.2cm}

\noindent and the dot multiplication are given as $M_i \cdot M_j= \varepsilon_{ab}M^a_iM^b_j$. Note that the SU(2k+1) invariant operators are defined by

\vspace{0.2cm}

$M^a_i \equiv {\bar Q^{\alpha}_i}Q^a_{\alpha}$ \hfill a, b=1, 2.

\vspace{0.2cm}

$X_{ij} \equiv A_{\alpha \beta} {\bar Q^{\alpha}_i} {\bar Q^{\beta}_j}$ \hfill i, j=1, $\cdots$, 2k-1.

\vspace{0.2cm}

$Y^a \equiv Q^a_{\alpha_{2k+1}} \varepsilon^{\alpha_1 \cdots \alpha_{2k+1}} A_{\alpha_1 \alpha_2} \cdots A_{\alpha_{2k-1} \alpha_{2k}}$ \hfill (13)

\vspace{0.4cm}

\noindent For $K \geq 2$,  the SU(2) moduli spaces are not modified quantum mechanically.  The SU(2k+1)$\times$SU(2)$\times$(1) invariant operators are subject to the constraints which follow from Bose statistics.  If we perturb the superpotential only by
 renormalizable operators at $\lambda_i \ll 1$, the most general superpotential should be:

\vspace{0.7cm}
$W_{eff} = W_{ren} + W_{dyn}$ \hfill (14)

\vspace{0.3cm}

$W_{ren} = \lambda^{ij}_1SL_{ij} + \lambda^{ij}_2X_{ij} + \lambda^{ij}_3N_{ij}$ \hfill (15)

\vspace{0.8cm}
$W_{dyn}$ = {\large ${\Lambda^{4k+3} \over W_{i_1}\varepsilon^{i_1\cdots i_{2k-1}}X_{i_2i_3}\cdots X_{i_{2k-2}i_{2k-1}}}$}. \hfill (16)

\vspace{0.5cm}

\noindent For simplicity, we could choose some special forms of the coupling matrices by

\vspace{0.2cm}

$$\lambda_1^{ij}=\cases{1&(i,j)=(1,2), (3,4) ..... (2k-3, 2k-2)\cr
	0&otherwise.\cr}$$	

\vspace{0.2cm}

\hspace{1cm}$$\lambda_2^{ij}=\cases{1&(i,j)=(2,3), (4,5) ..... (2k-4, 2k-3),(2k-2,1)\cr
	0&otherwise.\cr}$$	

\vspace{0.2cm}

\hspace{1.6cm} $\lambda_3$ : an unit matrix.  \hfill (17)

\vspace{0.4cm}

\noindent Note that $\lambda_1$ and $\lambda_2$ are antisymmetric. To see that there is no flat directions classically in this model, we can follow the similar argument as we have done in the section two.  The F-flat conditions are:

\vspace{0.4cm}

$\lambda_1^{ij}L_1 \cdot L_j=0$ \hfill (18.1)

\vspace{0.2cm}

$\lambda_2^{ij} {\bar Q}_i^{\alpha} {\bar Q}_j^{\beta}=0$ \hfill (18.2)

\vspace{0.2cm}

$\lambda_3^{ij}L_i^a{\bar Q}^{\alpha}_j =0$ \hfill (18.3)

\vspace{0.2cm}

$2\lambda_1^{ij}SL^b_j + \lambda_3^{ij} {\bar Q}^{\alpha}_j Q^b_{\alpha} =0$ \hfill (18.4)

\vspace{0.2cm}

$2\lambda_2^{ij}A_{\alpha \beta}{\bar Q}^{\beta}_j + \lambda_3^{ij}L_j^a Q^b_{\alpha} \varepsilon_{ab}=0$ \hfill (18.5) 
   
\vspace{0.4cm}

\noindent From equations (18.4) and (18.5), we get $M^b_{2k-1}=0=N_{i,2k-1}=W_{2k-1}=C_{i, 2k-1}=C_{2k-1,i}$ and $P_i=0=W_k$.  The relations between $N_{ij}$, $X_{ij}$ and $SL_{ij}$ can be obtained as

\vspace{0.4cm}

$N_{ik}=-2X_{kg(i)}$

\vspace{0.2cm}

$N_{ik}=-2SL_{if(k)}$ \hfill (19)

\vspace{0.4cm}

\noindent These relations then are used to determine the vevs of $N_{ik}$ as similar to determining $U_{ij}$ in section two.  Take 2k=10 as an example we have

\vspace{0.4cm}

$$N_{ik}=\left[\matrix{0&0&x&0&y&0&z&0&0   \cr
                       0&0&0&z&0&y&0&x&0    \cr
	               z&0&0&0&x&0&y&0&0    \cr	
                       0&x&0&0&0&z&0&y&0    \cr
                       y&0&z&0&0&0&x&0&0    \cr
	               0&y&0&x&0&0&0&z&0    \cr
                       x&0&y&0&z&0&0&0&0     \cr		 					
                       0&z&0&y&0&x&0&0&0    \cr
                       0&0&0&0&0&0&0&0&0    
 					}\right].$$ 

\vspace{0.4cm}

\noindent The vevs x, y and z can be shown to be zero to be consistent with constraint equations (18).  That is, all vevs of tensor operators $N_{ik}$, $X_{ik}$ and $SL_{ik}$ vanish. On the other hand, operators $SP_i$ and $W_k$ also have vanishing vevs. 
 We are left with operators $C_{ij}L_{kl}$.  By considering that all vevs of $N_{ik}=0$, the vevs of operators $L_i$ and $M_k$ must be one of the following cases.

\vspace{0.4cm}

\hspace{1cm} 1: All $M_k=0$.

\vspace{0.2cm}

\hspace{1cm} 2: All $L_k=0$.

\vspace{0.2cm}

\hspace{1cm} 3: $L_i$ // $M_k$ for those $L_i \not=0$ and $M_k \not= 0$. 

\vspace{0.4cm}

\noindent It is easy to see that all $C_{ij}L_{kl}$ vanish in cases 1, 2 and 3.  Thus we conclude that classically all vevs of gauge invariant operators vanish and there is no flat direction.    
From the discussion above we know that by adding appropriate renormalizable tree level superpotential terms to the effective superpotential, we can lift all classical flat directions. In this section, the pattern of tree level Yukawa couplings are the sam
e of those in the section 2. The SUSY critera guarantee that SUSY is spontaneously broken when the $U(1)_R$ symmetry is spontaneously broken by a dynamically generated non-perturbative superpotential term in this model.

For the case K=1, this becomes the well known SU(3)$\times$SU(2) model except for an additional singlet field S.  However, the U(1) charge of S hinders the interaction between S and other fields.  For K=2, it is the SU(5)$\times$SU(2)$\times$U(1) model di
scussed in [6].  

\vspace{1cm} 

\noindent {\Large 4. \hspace{1cm}SU(2k+2)$\times$U(1)}

\vspace{0.5cm}

This model has been well discussed in the literature [4b,6].  We  give just  a brief summary in this section. When the original SU(2k+3) theory breaks down to SU(2k+2)$\times$U(1), it leaves one antisymmetric tensor $(A_{\alpha \beta})_{-2}$, one fundamen
tal $Q_{\alpha}(2k+2,1)_{2k+1}$, 2k-1 antifundamentals ${\bar Q}^{\alpha}_i({\overline {2k+2}}, 1)_1$ and 2k-1 singlets $S_i(1,1)_{-(2k+2)}$ in the resulting theory. The U(1) coupling is presumed to be weak at the scale at which the SU(2) interaction beco
mes strong. Under the weak U(1) interaction assumption, the SU(2k+2) gauge dynamics generates a non-perturbative superpotential as the one in a SU(2k+2) theory.  

\vspace{0.3cm}

\hspace{1cm} $ W_{dyn}$ = {\large$ {\Lambda^{2k+3} \over { \lbrack A_{i_1} \varepsilon ^{i_1 \cdots i_{2k-1}} X_{i_2i_3} \cdots X_{i_{2k-2}i_{2k-1}} \rbrack }^{ 1 \over 2}}$}. \hfill (20)

\vspace{0.3cm}

\noindent where

\vspace{0.2cm}

\hspace{1cm} \makebox[4cm][l]{$A_i \equiv M_iPfA$}, \makebox[4cm][l]{$B_{ij} \equiv S_iM_j$.} \hfill $i,j=1, \cdots 2k-1.$

\vspace{0.2cm}

\noindent The singlets $S_i$ play a crucial role in breaking SUSY. The classical equation of motion of $S_i$
forces $A_i$ and $B_{ij}$ to have zero vev's and thus leads to no flat directions classically.  On the other hand,  the accidental non-anomalous $U(1)_R$ symmetry is lifted by the dynamically generated superpotential if the Yukawa coupling of $B_{ij}$ is 
a non-degenerate Yukawa matrix.  Thus according to the SUSY breaking criteria in section two, SUSY is dynamically broken in this model. 
      
\vspace{0.6cm}

\noindent {\Large 5. Conclusion }

\vspace{0.5cm}

The non-perturbatively generated effective superpotential as well as the quantum deformation of a moduli space can lead to supersymmetry breaking.  In this paper, we discuss three classes of models which can be constructed by decomposing a chiral gauge SU
(2k+3) theory with one antisymmetric tensor and 2k-1 antiquarks.  The original SU(2k+3) theory is well known to break SUSY once appropriate Yukawa terms are added to the effective superpotential.  By breaking the SU(2k+3) gauge theory down to the various 
theories with gauge groups as subgroups of SU(2k+3), we have resulting theories with zero Witten indices which thus make them good candidates for dynamical supersymmetry breaking.  The resulting SU(2k+1)$\times$SU(2)$\times$U(1) models break SUSY spontane
ously in the weak SU(2) limit for K$\geq$2 by the non-perturbative superpotential generated by the SU(2k+1) dynamics.  For K=1, this becomes the well known SU(3)$\times$SU(2) model which breaks supersymmetry either by the non-perturbative superpotential i
n the $\Lambda_2 \gg \Lambda_3$ limit or by the deformation of the quantum moduli space in the $\Lambda_3 \gg \Lambda_2$ limit.  The SU(2k)$\times$SU(3)$\times$U(1) models are discussed in the strong SU(2k) limit in this letter. By carefully choosing $E_{
ij}$ and $X_{ij}$ terms into the tree level superpotential, the theories can be constructed to be anomaly free and also respect the global $U(1)_R$ symmetry.  The $U(1)_R$ symmetry is then spontaneously broken by the deformation of the moduli space and th
us fulfill the SUSY-breaking criteria.  The SU(2k+2)$\times$U(1) model discussed in the section 4, it breaks SUSY by the non-perturbative superpotential generated by the SU(2k+2) dynamics.  The singlets fields $S_i$ play crucial roles in breaking SUSY.  T
hey lead to no flat direction classically and satisfy one of the two SUSY breaking criteria listed in the section three.    

During preparation of this letter, we found that Csaki et. al. [8] had finished work specially on the SU(2k)$\times$SU(3)$\times$U(1) model. 

\vspace{0.6cm}

\noindent {\Large 6. Acknowledgments }

\vspace{0.3cm}

We would like to thank M. Peskin, S. Thomas and Y. Shirman for helpful discussions.  We also thank A. Rajaraman for providing related information.  This work was supported by DOE grant $\sharp$ DE-AC03-76SF00515.

\vspace{1cm}

\newpage
\noindent {\Large \bf References}

\vspace{1cm}

\noindent [1] K. Intriligator and N. Seiberg,  {\bf hep-th/9509066}.

\vspace{0.8cm}
\noindent [2] K. Intriligator and S. Thomas, {\bf hep-th/9603158}.
 
\vspace{0.8cm}
\noindent [3] T. ter Veldhuis, {\em Phys. Lett.} {\bf B367}(1996)157.

\vspace{0.8cm}
\noindent [4a] L. Affleck, M. Dine and N. Seiberg, {\em Nucl. Phys.} {\bf B256}(1985)557; 

\noindent [4b] E. Poppitz and S. Trivedi, {\em Phys. Lett.} {\bf B365}(1996)125.

\vspace{0.8cm}
\noindent [5] E. Witten,  {\em Nucl. Phys.} {\bf B212}(1982)253.

\vspace{0.8cm}
\noindent [6] M. Dine, A. Nelson, Y. Nir and Y. Shirman, {\em Phys. Rev.} {\bf D53}(1996)2658.

\vspace{0.8cm}
\noindent [7] A. Nelson and N. Seiberg, {\em Nucl. Phys.} {\bf B416}(1994)46.

\vspace{0.8cm}
\noindent [8] C. Csaki, L. Randall and W. Skiba,  {\bf hep-th/9605108}.

\end{document}